\documentstyle[prl,aps]{revtex}
\input epsf
\begin{document}
\draft

\twocolumn[\hsize\textwidth\columnwidth\hsize
\csname@twocolumnfalse\endcsname

\title{Chaos and Universality in the Dynamics of Inflationary
Cosmologies} 

\author{H. P. de Oliveira\thanks{E-mail: oliveira@dft.if.uerj.br}, S. L.
Sautu\thanks{E-mail:ona@dft.if.uerj.br}}

\address{{\it Universidade do Estado do Rio de Janeiro }\\ {\it
Instituto de F\'{\i}sica - Departamento de F\'{\i}sica Te\'orica,}\\
{\it CEP 20550-013 Rio de Janeiro, RJ, Brazil.}}

\author{I. Dami\~ao Soares\thanks{E-mail: ivano@lca1.drp.cbpf.br} and E.
V. Tonini\thanks{E-mail: tonini@lca1.drp.cbpf.br}} 

\address{{\it Centro Brasileiro de Pesquisas F\'\i sicas\\ Rua Dr.
Xavier Sigaud, 150 \\ CEP 22290-180, Rio de Janeiro, RJ, Brazil.}}

\maketitle 


\begin{abstract}
We describe a new statistical pattern in the chaotic dynamics of closed
inflationary cosmologies, associated with the partition of the
Hamiltonian into rotational motion energy and hyperbolic motion energy
pieces, in a linear neighborhood of the saddle-center present in the
phase space of the models. The hyperbolic energy of orbits visiting a
neighborhood of the saddle-center has a random distribution with respect
to the ensemble of initial conditions, but the associated histograms
define a statistical distribution law of the form $p(x) =
C\,x^{-\gamma}$, for almost the whole range of hyperbolic energies
considered. We present numerical evidence that $\gamma$ determines the
dimension of the fractal basin boundaries in the ensemble of initial
conditions. This distribution is universal in the sense that it does not
depend on the parameters of the models and is scale invariant. We
discuss possible physical consequences of this universality for the
physics of inflation.
\end{abstract}

\vskip2pc]

The general dynamics of closed Friedmann-Robertson-Walker (FRW)
inflationary cosmologies, with a minimally coupled massive scalar field
and perfect fluid or with anisotropy, is chaotic\cite{mos,oss}. The
chaotic behaviour has a definite homoclinic origin, due to infinite
transversal crossings of homoclinic cylinders present in the phase space
of the models, analogous to the case of breaking and crossing of
homoclinic connections in Poincar\'e's homoclinic
phenomena\cite{wiggins,vieira,holmes1,holmes,moser}. The topology of
homoclinic cylinders is a consequence of a critical point of the
saddle-center type\cite{holmes} in phase space, and the breaking and
crossings of the cylinders are due to the non-integrability of the
dynamics. These homoclinic phenomena produce chaotic sets in phase
space, in the sense that the dynamics is highly sensitive to
fluctuations in initial conditions taken on these sets. The aim of this
communication is to describe a new statistical pattern in the chaotic
dynamics engendered by the saddle center and what are its possible
consequences in the physics of inflation. Throughout the paper we use
units such that $8\,\pi\,G = 1 = c$.

Our discussion will be basically restricted to closed
Friedmann-Robertson-Walker (FRW) inflationary cosmologies, with a
minimally coupled massive scalar field and perfect fluid, but the same
construction will apply to a Bianchi IX anisotropic cosmology and yields
analogous results, as we will see. The equations of motion may be
derived from the Hamiltonian constraint

\begin{equation}
H = -\frac{p_{\varphi}^2}{2\,a^2} + \frac{p_a^2}{12} + 3\,a^2 -
a^4\,V(\varphi) - E_0 = 0, \label{eq.1}
\end{equation}

\noindent where $a$ and $\varphi$ are the scale factor and the scalar
field, respectively, and $p_a$ and $p_{\varphi}$ are the momenta
canonically conjugate to $a$ and $\varphi$. We consider here a perfect
fluid describing radiation, and the potential $V(\varphi) =
\Lambda+\frac{1}{2}\,m^2\,\varphi^2$. The constant $\Lambda$ is
considered to correspond to the vacuum energy of the scalar field
$\varphi$, and plays the role of a cosmological constant term. This
model may be thought to describe a classical universe just exiting a
Planck era. $E_0$ is a constant arising from the first integral of the
Bianchi identities, and proportional to the total matter energy of the
models. The dynamical system generated by (1) has a saddle-center
critical point in the finite region of the phase space with coordinates
$E: \varphi=0, a=a_0=\sqrt{\frac{3}{2\,\Lambda}}, p_a=0, p_{\varphi}=0$.
The energy associated to the critical point $E$ is
$E_0=E_{crit}=\frac{3\,a_0^2}{2}$. The phase space also has a pair of
critical points at infinity, corresponding to the DeSitter solution, one
acting as an attractor, and the other as a repeller; the DeSitter
attractor characterizes an exit to inflation for the models. 

In accord to a theorem by Moser\cite{moser1}, it is always possible to
find a set of canonical variables such that, in a linear neighborhood of
a saddle-center, the Hamiltonian dynamics is separable into rotational
motion and hyperbolic motion pieces, with the associated partial
energies $E_{rot}$ and $E_{hyp}$ approximately conserved. The canonical
variables $(a, \varphi, p_a, p_{\varphi})$ behave as Moser's variables
in a small neighborhood of $E$; in fact, expanding the Hamiltonian (1)
about $E$ we obtain

\begin{equation}
H = E_{hyp} - E_{rot} + (E_{cr} - E_0) + {\cal{O}}(3) = 0, \label{H1}
\end{equation}

\noindent where $E_{rot} =
\frac{1}{2}\,\frac{p_\varphi^2}{a_0^2}+\frac{1}{2}\,m^2\,a_0^4\,\delta\,
\varphi^2$ and $E_{hyp} =\frac{1}{12}\,p_a^2 - 6\,\delta\,a^2$, with
$\delta\,a = a - a_0$, $\delta\,\varphi = \varphi$, and $(E_{cr}-E_0)$
small. ${\cal{O}}(3)$ denotes higher order terms which are neglected in
a small neighborhood of $E$. In this approximation the scale factor
$a(t)$ has pure hyperbolic motion, and is completely decoupled from the
scalar field pure rotational motion in the plane $(\varphi,
p_{\varphi})$.

We will now focus our attention to chaotic aspects in the dynamics,
leading us into a statistical treatment of possible experiments made in
the models. Our description will be restricted to sets of orbits which
visit a linear neighborhood of the saddle-center $E$, and makes
extensive use of the partition of energy $E_{hyp} - E_{rot} + (E_{crit}
- E_0) \approx 0$ that occurs in this neighborhood. Let us consider a
set ${\cal{D}}$ of arbitrary initial conditions which correspond to
initially expanding universes with energy $E_0$, and such that all the
orbits visit a linear neighborhood of $E$. The set ${\cal{D}}$ may be
represented as a small 4-dim volume, contained in a 4-dim sphere of
radius $R$ (of order $10^{-5}$, $10^{-6}$, ..., depending on the order
of fluctuations we admit in the initial conditions) constructed about
the point, say, $(0.4, 0, 3.5638181771, 0)$ in phase space, and
satisfying the Hamiltonian constraint (1). The above point belongs to
the invariant plane $p_\varphi=\varphi=0$, and is located on one of the
separatrices reaching the saddle-center $E$. We denote $R$ as the
characteristic radius of the four volume ${\cal{D}}$. One of the chaotic
aspects we single out for this set is the definition of a chaotic exit
to inflation\cite{mos,oss}, connected to the existence of {\it fractal
basin boundaries} in the set ${\cal{D}}$. For a given radius $R$ of
${\cal{D}}$, it is always possible to find a gap of energy $\delta\,E^*$
(which depends on $R^2$) such that for $E_0 \in \delta\,E^*$, the exit
to inflation is chaotic, namely, small fluctuations (of order of, or
smaller than $R$) of the initial conditions will change the asymptotic
state of the orbits from collapse into escape to inflation, and vice
versa, after visiting the neighborhood of the saddle-center. The chaotic
exit to inflation may be characterized in terms of the partition of
energy (2): for a particular orbit, the partition into the approximately
conserved energies $E_{hyp}$ and $E_{rot}$ about the critical point will
determine the asymptotic state of the orbit, namely, collapse or escape
to inflation, whether respectively $E_{hyp} < 0$ or $E_{hyp} > 0$.
However, we are no longer able to foretell precisely what will be the
value of $E_{hyp}$ endowed to a given orbit of ${\cal{D}}$ when it
visits a neighborhood of $E$, due to the high sensitivity of $E_{hyp}$
with the fluctuations of the initial conditions, a consequence of the
chaotic nature of the set ${\cal{D}}$. In what follows, we will use
standard statistical methods to examine this instability of the
partition of the energy, a procedure which will lead us into the
recognition of new statistical patterns for chaotic sets in the phase
space.

To start, let us make a plot of $E_{hyp}$ versus the ordered values of
one of the canonical variables, say $a$, especifying the initial
conditions in the set ${\cal{D}}$, for $R \approx 10^{-5}$, $m=2$ and
$(E_{crit} - E_0) = 10^{-10}$. We fixed the value of $\Lambda=3/2$, such
that $a_0=1$ and $E_{crit}=3/2$. Fig. 1 corresponds to $3,031$ initial
conditions, sampled randomly and uniformly in ${\cal{D}}$, and the
points were connected for successive a's to a better visualization. The
result shows a highly disorganized "signal", displaying structures in
all scales of $a$. Analogous highly disorganized "signals" appears for
other variables $\varphi$, $p_a$, or $p_\varphi$. Also, if another
sample of points is taken randomly and uniformly in ${\cal{D}}$, we
obtain a "signal" with the same general aspect but completely different
in all the details, appearing not to have any connection with the
previous figure. However a pattern of the signal is reproducible, which
is its histogram. 

Our basic procedure is to realize the following numerical experiments
with the system: from a given ensemble ${\cal{D}}$ of initial
conditions, we generate orbits of the system by the full Hamiltonian
dynamical equations derived from (1). These orbits visit a linear
neighborhood of the saddle-center $E$. In this linear neighborhood, we
calculate the approximately conserved quantity $E_{hyp}$ associated with
each orbit ; for practical purposes, $E_{hyp}$ is evaluated at the point
of closest approach to $E$, in the sense of the Euclidean measure. We
construct the histogram function
$P_{E_{hyp}}=\Delta\,N(E_{hyp})/\Delta\,E_{hyp}$ as usual: the $E_{hyp}$
axis is divided into a large number of equal small bins
$\Delta\,E_{hyp}$ and the histogram is given by the number of orbits
assigned with hyperbolic energy in the interval
$[E^{(i)}_{hyp}-\Delta\,E_{hyp}/2, E^{(i)}_{hyp}+\Delta\,E_{hyp}/2]$ ,
when the orbit visits the neighborhood of $E$. Here $E^{(i)}_{hyp}$ is
the value of the hyperbolic energy centered in the $i$-th bin. We fix
the origin of the histogram in $E^{min}_{hyp}$, which is the minimum
value of $E_{hyp}$ obtained in the numerical experiment; according to
(2), $E^{min}_{hyp} \geq - (E_{crit}-E_0)$. We also note that due to (2)
the histogram of the signal associated with $E_{rot}$ results merely
from a shift of the histogram for $E_{hyp}$.

A careful examination shows us that the form of the histogram for
smaller values of $E_{hyp}$ is sensitive to the geometry of the sampling
set ${\cal{D}}$. We must therefore establish a criterion to fix the
geometry of ${\cal{D}}$ such that no bias due to the geometry of the
sampling volume manifests itself in the histogram, and such that the
statistical distribution function obtained from the histogram will be
connected only to the non-integrability of the system. We establish that
the domain ${\cal{D}}$ of randomly and uniformly distributed phase space
points must have a geometry such that it yields histograms
$\Delta\,N(E_{hyp})/\Delta\,E_{hyp}$ or
$\Delta\,N(E_{rot})/\Delta\,E_{rot}$ for the integrable Hamiltonian $H =
\frac{1}{2}\,(p_{\varphi}^2+m^2\,\varphi^2) +
\frac{1}{12}\,p_a^2-6\,a^2$ having the form of a plateau (cf. also the
caption of Figs. 2). In the case of the non-integrable Hamiltonian (1)
the histogram constructed from the {\it same} set of initial conditions
${\cal{D}}$ deviates from a plateau, reflecting the nonintegrability of
the system. Figs. 2(a,b) show $30,000$ initial conditions sampled
randomly and uniformly in ${\cal{D}}$ constructed for $m = 10$, $R
\approx 10^{-5}$ and $(E_{crit} - E_0) = 10^{-10}$, projected on the
planes $(\varphi,p_{\varphi})$ and $(a,p_a)$. Fig. 3 shows the histogram
generated from the initial conditions set ${\cal{D}}$ of Fig. 2 by the
non-integrable Hamiltonian (1).

At this point our main question is: what are the possible candidates for
the statistical energy distribution law associated with the histogram
function? The present histograms have an excess in the tails, namely, in
the region of large positive hyperbolic energies, relative to the
corresponding Maxwell-Boltzmann distribution. Indeed, a trivial
log-linear plot of the histogram shows a non linear form for the tails,
exhibiting an anomalous behaviour in contrast to the case of the normal
Maxwell-Boltzmann distribution. This is not surprizing since we are
dealing with a system were long range interactions are present, and
where relevant phase-space sets are chaotic and have a fractal-like
structure. The system is certainly not in the realm of Boltzmann-Gibbs
statistical mechanics, which essentially focus on non-interacting or
short-range interacting many-body Hamiltonian systems. After extensive
numerical analysis, we obtained that all histograms examined are fitted
by the energy distribution function 

\begin{equation}
P(x) =  C_0\,\beta\,(\beta\,x)^{-\gamma},
\end{equation}

\noindent where $C_0$ is an adimensional normalization constant and $x =
(E_{hyp}-E^{min}_{hyp})$. The parameters of the distribution have values
$\gamma = 0.6354 \pm 0.02527$,
$C_0\,\beta^{1-\gamma}\,\times\,10^{10\,\gamma} = 12.496 \pm 0.5108$.
These are the values for the best fit of the histogram of Fig. 3, with
the square of the variance $\chi^2 = 0.06022$, obtained using the PAW
packet from CERN. Within statistical errors, the value $\gamma \cong
0.6354$ corresponds to the best fit of {\it all} histograms generated
with the dynamics of (1). For comparison purposes, all histograms were
normalized by fixing the peak with the value $100$. 

The distribution law (3) produces an accurate fit for almost the whole
range of energies of the histograms. The continuous line in Fig. 3 shows
the best fit of (3) to the histogram. The whole histogram is composed of
$100$ bins, each with $\Delta\,E_{hyp} = 7.33993555\times10^{-12}$. To
obtain the fit with the value $\chi^2 = 0.06022$, we have neglected the
extreme of the tail of the histogram, more specifically from the
$55^{th}$ bin on. If we have extended the solid curve beyond this bin, a
discrepancy between the curve and the histogram would appear. This
discrepancy at the end of the tail has the following explanation. The
cut-off at the largest values of $E_{hyp}$ is due to the finite volume
of the sampling set ${\cal{D}}$. If we increase $R$ (consequently
increasing the volume) and maintain the same uniform density of points,
the cut-off is shifted to the right and the tail of the histogram raises
improving the fit in this region. The left extremity is a peak. If we
extended (3) to values of $x$ very close to zero, the peak might suggest
a divergence in the density of states - a property which is not
physically reasonable for statistical ensembles - but this is not the
case. Indeed the smallest values of $E_{hyp}$ in our histograms are
associated to points that are located at the boundaries in $(a, p_a)$ of
the linear neighborhood of the saddle-center $E$, and the evidence from
the histograms is that, as we increase the volume of ${\cal{D}}$, there
is a saturation in the corresponding increase of the peak of the
histograms. Therefore the distribution function (3) must surely be
corrected, to be finite in the neighborhood of $x = 0$. The saturation
is possibly due to the fact that, as we enlarge the volume, we approach
the limit of validity of the linear approximation, and higher order
terms in the Hamiltonian may be contributing already to make the density
of states finite there. The parameter $\gamma$ is adimensional while
$1/\beta$ must have the dimension of energy. We may tentatively
interpret the latter as defining a {\it temperature} for the
saddle-center. For the integrable case, $\gamma = 0$. 

The following characterist properties of the histograms were verified
numerically, through exhaustive numerical experiments realized with
randomly and uniformly distributed distinct sampling sets ${\cal{D}}$
having the same geometry described in Figs. 2:

\noindent $(i)$ the form of the histogram is scale invariant, that is,
it is independent of the characteristic radius $R$ of the sampling set
${\cal{D}}$ of initial conditions, modulo a change of bin;

\noindent $(ii)$ the form of the histogram is independent of the
parameter $(E_{crit}-E_0)$ that allows to characterize if the orbits are
dominantly collapsing or escaping, after visiting the neighborhood of
the saddle-center $E$;

\noindent $(iii)$ the form of the histogram is independent of the mass
parameter $m$;

\noindent $(iv)$ the histogram has the same form also for a model with
dust, what lead us to conjecture its independence of the particular
equation of state $p = \lambda\,\rho$, with $0 \leq \lambda \leq 1$;

\noindent $(v)$ within statistical errors, the value $\gamma \cong
0.6354$ corresponds to the best fit of all histograms associated with
the Hamiltonian (1).

We were not able to identify (3) with any known statistical
distribution. However, we are aware that this behaviour is typical of
tails of distributions arising in Tsallis generalized
statistics\cite{tsallis}, that were proposed in order to accomodate, at
least in part, systems which have an anomalous behaviour with respect to
Boltzmann-Gibbs statistics.
 
We now exhibit numerical evidence that the constant $\gamma$ appearing
in (3) may characterize the fractal nature of the sets ${\cal{D}}$, more
specifically, it may give a measure of the fractal dimension of basin
boundaries in these sets. In fact, by using a box counting
method\cite{ott} with the uncertainty code defined by collapse/escape,
we obtained the uncertainty exponent $\alpha \cong 0.36931$ for the set
${\cal{D}}$ with $m = 10$, $(E_{crit} - E_0) = 10^{-10}$ and $R =
10^{-5}$ (cf. Fig. 5). Within statistical errors, this same value was
obtained for all initial conditions sets with $E_0 \in \delta\,E^*$,
used to construct histograms in our numerical experiments, implying that

\begin{equation}
\gamma+\alpha \approx 1.
\end{equation}

\noindent As a consequence $\gamma$ appears to be directly related to
the fractal dimension $d$ of the basin boundaries of ${\cal{D}}$, since
in our boxcounting calculation $d = N - \alpha$, where $N = 3$ is the
dimensionality of ${\cal{D}}$. The above restriction to sets ${\cal{D}}$
with $E_0 \in \delta\,E*$ is due to the fact that in these sets we may
define the uncertainty code collapse/escape; for sets with $E_0 \notin
\delta\,E^*$, other uncertainty code may be used. 
 
We have also examined the Hamiltonian

\begin{eqnarray}
H & = & \frac{1}{4\,B}\,P_A\,P_B - \frac{1}{8\,B^2}\,A\,P_A^2 + 2\,A -
\frac{1}{2}A^3\,B^2 - \nonumber \\ & & 2\,\Lambda\,B^2 - E_0 =0, 
\end{eqnarray}

\noindent describing the dynamics of anisotropic spatially homogeneous
Bianchi IX cosmological models, with scale factors $A$ and $B$, plus a
cosmological constant and a perfect fluid in the form of dust. This
system is chaotic and has a saddle-center in the finite region of phase
space\cite{oss}. We made the same construction of histograms for
hyperbolic/rotational energies in a linear neighborhood of the
saddle-center; in this case, we have the burden of obtaining Moser's
normal coordinates in order to calculate properly the
hyperbolic/rotational energies associated to each orbit when visiting
the neighborhood of the saddle-center. Fig. 4 shows the histogram
generated from 30,000 initial conditions. The geometry of the initial
conditions set satisfies the same criteria of the previous case (1). The
continuous line corresponds to the best fit of (3), with $\gamma =
0.71458 \pm 0.04060$, $C_0\,\beta^{1-\gamma}\times10^{10\gamma} = 11.601
\pm 0.7711$, and $\chi^2 = 0.06685$. The histograms consists of 65 bins,
each with value $\Delta\,E_{hyp} = 3.58919436\times10^{-12}$. The best
fit was realized in the interval of the $2^{nd}$ to the $30^{th}$ bin.
Within statistical errors, all histograms associated with (5) are fitted
by the distribution function (3) with $\gamma \cong 0.714$. Properties
$(i)$ and $(ii)$ were also verified for the present case. A boxcounting
for the initial conditions set yields the uncertainty exponent $\alpha
\cong 0.28103$, confirming the relation $\gamma + \alpha \cong 1$.

We end this paper by making the conjecture that the statistical
distribution (3) is an universal characteristic of saddle-center
critical points, independent of the particular dynamical system in
consideration. We are presently working in this direction by making the
same {\it constructo} for two other Hamiltonian systems presenting a
saddle-center in their phase space, not in the realm of inflationary
cosmologies: the H\'enon-Heiles system\cite{henonheiles} and the
chemical system examined in Ref. \cite{osorio}. Our preliminary results
indicate the same distribution function (3) with distinct values of
$\gamma$ for each system. We also conjecture whether other critical
points of different nature can be universally characterized in this way.

The above results may have some interesting physical applications for
the dynamics of preinflationary models immediately before the universe
enters the inflationary regime or recollapses, both alternatives
depending on the typical fluctuations in the system. For instance, we
may use the statistical function (3) to calculate the average value of
the hyperbolic energy for ensembles of universes. As we
know\cite{mos,oss} the sign of this average hyperbolic energy determines
the "average" long time behaviour of the ensemble, that is, collapse or
escape to inflation. A straightforward calculation shows that the sign
of the average energy depends basically on $((1-
\gamma)\,\beta\,E^{max}_{hyp} + \beta\,E^{min}_{hyp})$, where
$E^{max}_{hyp}$ is the maximum value of the hyperbolic energy in the
histogram. The relative value of $E^{min}_{hyp}$ decreases linearly (and
may exceed in absolute value $E^{max}_{hyp}$) with the increase of
$(E_{crit} - E_0)$, implying that larger/smaller values of $(E_{crit} -
E_0)$ determine the dominance of collapse/escape in the dynamics of the
ensemble ${\cal{D}}$ of universes.

\noindent The authors are grateful to Prof. C. Tsallis for stimulating
discussions concerning the generalized statistics. We are also indebted
to the graduate students of LAFEX/CBPF, for introducing us to the use of
PAW. Financial support from CNPQ and FAPERJ is acknowledged. H. P. de
Oliveira is grateful to the International Center of Theoretical Physics
(ICTP) for support.

\newpage

\vspace{-5cm}

\begin{figure}[t]
\begin{minipage}[t]{0.49\textwidth}
\epsfysize=11.5cm
\epsffile{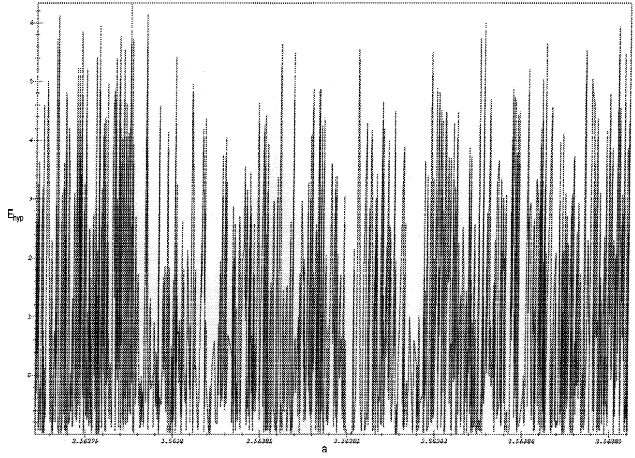}
\vspace{-2cm}
\caption{Signal resulting from the distribution of hyperbolic
energies $E_{hyp}$ versus the ordered initial values of $a$. We
considered 1,035 initial conditions sampled randomly and uniformly.
Points were connected for successive $a$'s, for better visualization.}
\end{minipage}
\end{figure}

\vspace{-3.7cm}
\begin{figure}[t]
\begin{minipage}[t]{0.49\textwidth}
\epsfysize=13cm
\hspace{-1cm}
\epsffile{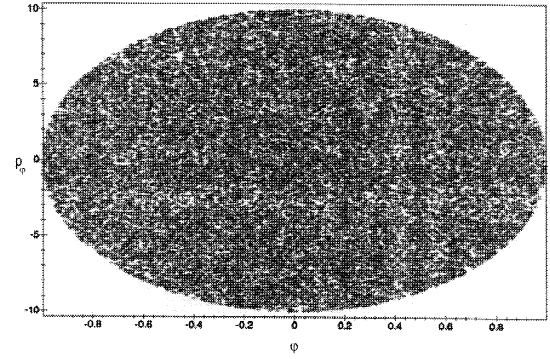}
\end{minipage}
\begin{minipage}[t]{0.49\textwidth}
\vspace{-8cm}
\epsfysize=13cm
\hspace{-1cm}
\epsffile{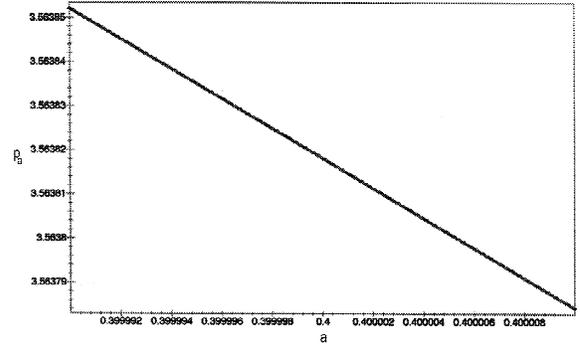}
\vspace{-4cm}
\caption{Set ${\cal{D}}$ of 30,000 initial conditions,
uniformly and randomly distributed, for the case $m=10$, $E_{crit} -
E_0=10^{-10}$ and $R = 10^{-5}$, projected in the planes
$\varphi,p_{\varphi}$ (Fig. 2a) and $(a,p_a) (Fig. 2b)$. The particular
geometry taken for the initial conditions set is such that $(i)$ it
satisfies the Hamiltonian constraint (1) and, $(ii)$ the associated
histograms for the integrable case have a form of a plateau; this
demands that Fig. 2a be an ellipse with axes having the ratio $m$.}
\end{minipage}
\end{figure}
\hspace{0cm}


\vspace{-5cm}
\begin{figure}[t]
\begin{minipage}[t]{0.49\textwidth}
\epsfysize=13cm
\hspace{-1cm}
\epsffile{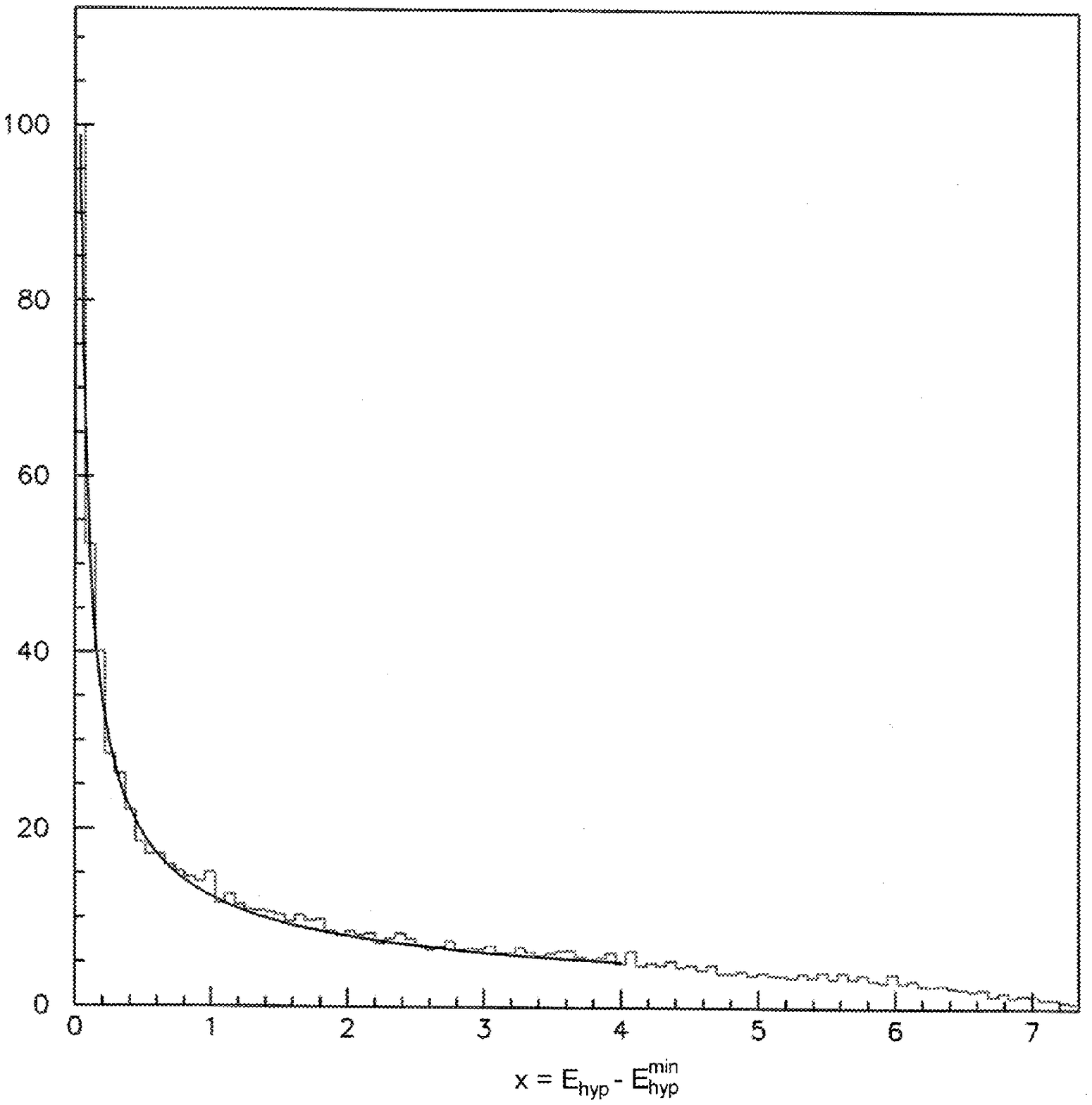}
\vspace{-2.5cm}
\caption{Histogram of the distribution of $E_{hyp}$, associated
with the 30,000 initial conditions of Fig. 2. The continuous line is the
best fit of the distribution (3) with $\gamma = 0.6354 \pm 0.02527$,
$C_0\,\beta^{1-\gamma}\times10^{10\gamma} = 12.496 \pm 0.5108$ and
$\chi^2 = 0.06022$. The horizontal axis is scaled with the factor 
$10^{-10}$.}
\end{minipage}
\hspace{0cm}

\begin{minipage}[t]{0.49\textwidth}
\epsfysize=13cm
\hspace{-1cm}
\epsffile{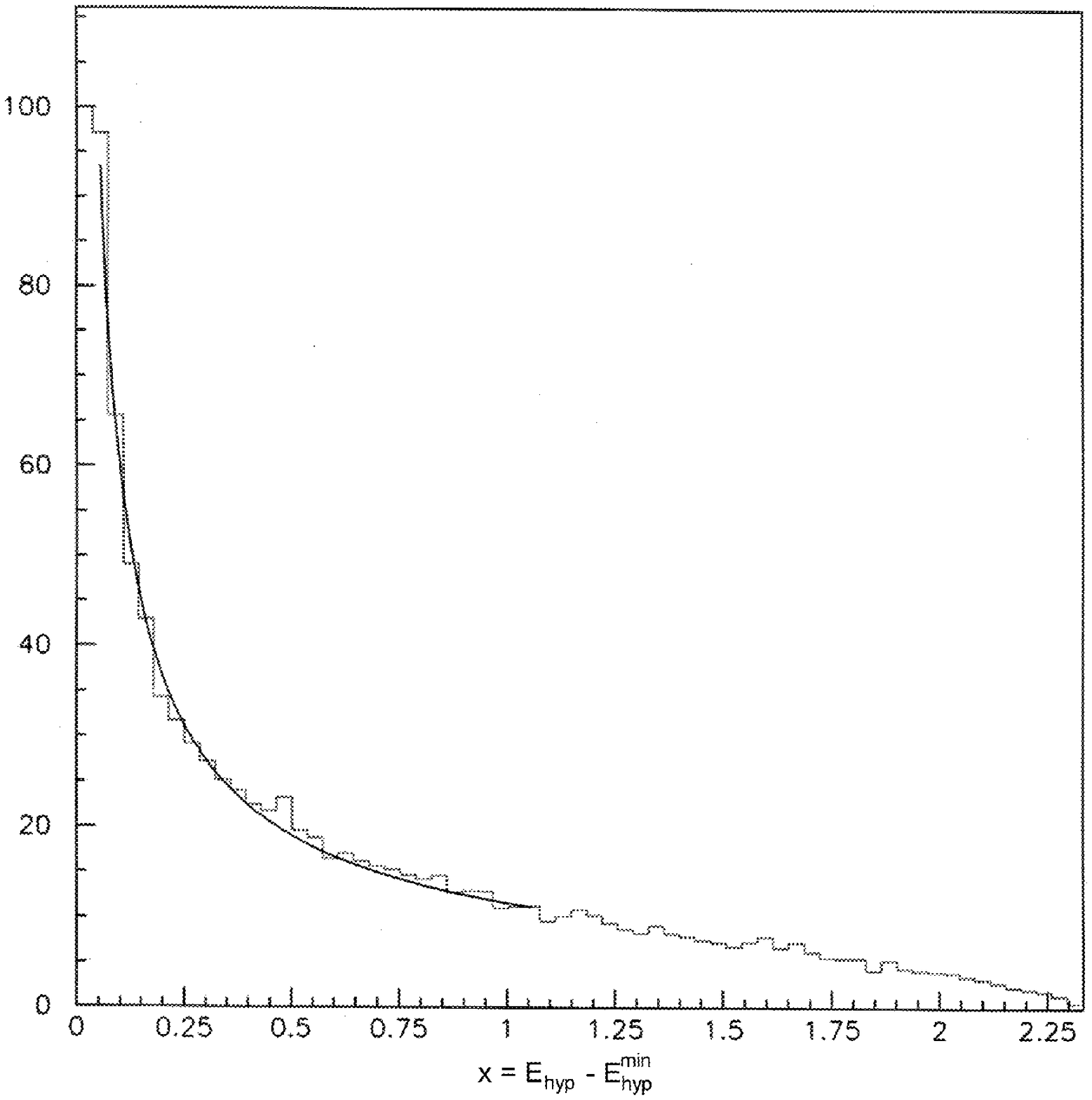}
\vspace{-2.2cm}
\caption{Histogram of the distribution of $E_{hyp}$, generated
from 30,000 initial conditions for the anisotropic case, with $\Lambda =
1/4$, $E_{crit} - E_0 = 0.5\times10^{-10}$ and $R = 10^{-5}$. The
continuous line is the best fit of (3) with $\gamma = 0.7146 \pm
0.0406$, $C_0\,\beta^{1-\gamma}\times10^{10\gamma} = 11.601 \pm 0.7711$
and $\chi^2 = 0.06685$. The horizontal axis is scaled with the factor 
$10^{-10}$.}
\end{minipage}

\end{figure}

\vspace{-4.3cm}
\begin{figure}
\begin{minipage}[htb]{0.49\textwidth}
\epsfysize=13cm
\hspace{-1.3cm}
\epsffile{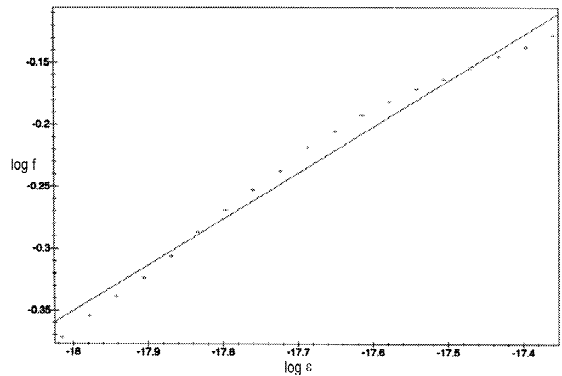}
\vspace{-3.8cm}
\caption{Plot of $f = k_0\,\epsilon^\alpha$ in a log-log scale,
where $\epsilon$ is the uncertainty radius about 20,000 initial
conditions taken inside the set ${\cal{D}}$, and $f$ is the fraction of
${\cal{D}}$ of uncertain initial conditions with the uncertainty code
collapse/escape. We considered 20 values of $\epsilon$ inside the
interval $[0.9 \times 10^{-8},1.8 \times 10^{-8}]$. The best fit renders 
$\alpha \approx 0.36931$.}

\end{minipage}
\end{figure}

\end{document}